\documentstyle[twoside,fleqn,npb,epsfig]{article}
\newcommand{\AmS}{{\protect\the\textfont2
  A\kern-.1667em\lower.5ex\hbox{M}\kern-.125emS}}
\newcommand{\PO}{I\!\!P}

\newcommand{\xpom}{x_{\PO}}
\newcommand{\ypom}{y_{\PO}}

\title{
\begin{flushright}\normalsize
  \vspace{-4cm}
\end{flushright}
High-$t$ Diffraction at HERA
}

\author{B. ~E. ~Cox\address{Department of Physics and Astronomy,
        University of Manchester, Brunswick Street, Manchester M13 9PL. UK}%
}
\begin{document}

\thispagestyle{empty}

\begin{abstract}
The double dissociation photoproduction cross section for the process 
$\gamma p \rightarrow XY$, in which the systems $X$ and $Y$ are separated by a large rapidity gap, is measured at large 4-momentum transfer squared $|t| > 20 \rm{GeV}^2 $ by the H1 Collaboration at HERA. This measurement provides for the first time a direct measurement of the energy dependence of the gap production process at high $|t|$.

\end{abstract}

\maketitle

\section{Introduction}
It is now an established experimental fact that there are events with large rapidity gaps in the hadronic final state in which there is a large momentum transfer across the gap. Such events have been observed at both the TEVATRON \cite{DZero,CDF} and HERA \cite{zeus,H1} in the rapidity gaps between jets process suggested for study by Bjorken \cite{Bj}. The issue now for experimentalists and theorists alike is to address the question of what underlying dynamical process is responsible for such striking events. It is clear that conventional Regge phenomenology cannot provide an answer, since the soft pomeron contribution has died away at much lower $|t|$ values due to shrinkage. The two best developed models currently available are the BFKL pomeron \cite{BFKL}, calculated within the leading logarithmic approximation (LLA) by Mueller and Tang \cite{MT} and implemented into the HERWIG Monte Carlo \cite{HERWIG,MH}, and the soft colour rearrangement model \cite{soft}. The recent gaps between jets analysis by the D0 Collaboration \cite{DZero} favoured the soft colour model to the BFKL pomeron, although conclusions from gaps between jets measurements may be difficult to draw due to the uncertainties in the role of multiple interactions, which are poorly understood theoretically at the present time \cite{JF,CFL}. Furthermore, gaps between jets measurements at both HERA and the TEVATRON are limited by the requirement that two jets are observed in the detector, severely restricting the accessible gap size. Since the BFKL cross section is predicted to rise exponentially with $\Delta \eta$, whilst soft colour is not, this is a severe restriction. At HERA, measurements of high $|t|$ vector meson production ~\cite{mesonexp,JC} have provided access to larger rapidity gaps in a well defined kinematic range, although the rate is low. With these issues in mind, Cox and Forshaw \cite{FC} suggested the study of the more inclusive double dissociative process $\gamma p \rightarrow XY$ at high $|t|$. In this paper we report the first measurement of this process, based on H1 data taken during 1996.

\begin{figure} 
\centerline{\epsfig{file=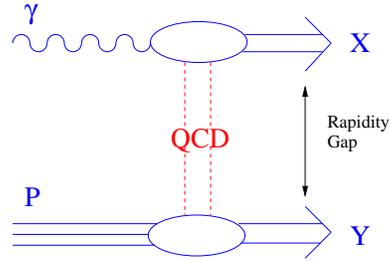,height=5.cm,angle=270}}
\caption{The double diffraction dissociation process at HERA.}
\label{diffplot}
\end{figure}

The photon and proton dissociative systems, $X$ and $Y$ respectively, are separated by finding the largest rapidity gap in the event (the procedure used by the H1 Collaboration in previous diffractive measurements \cite{H1diff}). 
The process, shown schematically in figure \ref{diffplot}, is considered in terms of the kinematic variables 
\begin{eqnarray}
W^{2}=(\gamma + P)^{2} \hspace{.5cm}
t=(P-Y)^{2} 
\end{eqnarray}
\begin{eqnarray}
\xpom=\frac{\gamma.(P-Y)}{\gamma.P} \hspace{.5cm}  
\ypom=\frac{P.(\gamma-X)}{\gamma.P}
\end{eqnarray} 
where $\gamma, P, X$ and $Y$ are the 4-vectors of the photon, proton and X and Y systems respectively. $W$ is the $\gamma P$ center of mass energy and $t$ is the four momentum transfer across the rapidity gap.
In this study we present measurements of the differential cross section ${\rm{d}} \sigma / {\rm{d}} \xpom (\gamma P \rightarrow XY)$ in the range $|t| > 20 {\rm GeV}^2$, $165 < W < 233 \ {\rm GeV}$, $0.0007 < \xpom < 0.0040$, $ \ypom < 0.018$.

\section{Event Selection}

The data for this analysis were collected with the H1 detector during the 
1996 running period, when HERA collided $27.6~{\rm GeV}$ positrons with $820~{\rm GeV}$ 
protons, with an integrated luminosity of 6.7 ${\rm pb}^{-1}$.
Photoproduction events were selected by detecting the scattered positron 
in the electron tagger, 33m down the beam pipe in the scattered electron direction. This 
restricts the virtuality of the photon to $Q^{2} < 0.01 $ GeV$^{2}$.

The reconstruction of the $X$ and $Y$ system 4-vectors has been optimised by combining tracking and calorimeter information. Techniques are applied to minimise the effects of detector noise. Precise details can be found elsewhere \cite{H1diff}. Losses in the forward and backward directions are, however, unavoidable, making the measurement of the invariant masses of the systems problematic. For this reason, we introduce the kinematic variables $\xpom$ and $\ypom$, reconstructed using the expressions

\begin{eqnarray}
\xpom \simeq \frac{M_{X}^{2}-t}{W^2} \simeq \frac{\Sigma{(E+p_z)_{X}}}{2E_p} \\
\ypom \simeq  \frac{M_{Y}^{2}-t}{W^2} \simeq \frac{\Sigma{(E-p_z)_{Y}}}{2E_\gamma}
\end{eqnarray}
where $E_p$ and $E_\gamma$ are the proton and photon beam energies respectively, and the quantity $\Sigma{(E+p_z)_{X}}$ ($\Sigma{(E-p_z)_{Y}}$) is summed over all hadrons reconstructed backward (forward) of the largest rapidity gap in the event. This quantity has the property that it is insensitive to losses down the beam pipe, for which $E \simeq -p_z$ ($E \simeq +p_z$).

In order to ensure that the systems $X$ and $Y$ are clearly separated, only events with a rapidity gap between the two systems of at least 1.5 units of rapidity are selected. These events are specified by $M_X, M_Y \ll W$, and hence our sample is defined in the kinematic range $\ypom < 0.018$ and $0.0007 < \xpom < 0.0040$ \footnote{Note that the requirement that the rapidity separation between the $X$ and $Y$ systems must be $>1.5$ is NOT part of the hadron level cross section definition. Any losses due to this cut are included in the acceptance corrections}.

The reconstruction of $t$ is more problematic. It is measured as the negative squared transverse momentum of the $X$ system, $-{(p_t)}_X$, and is sensitive to losses down the backward beam pipe, particularly for low values of $|t|$. For this reason we choose to define our sample for $|t| > 20 \rm GeV^2$.   

\section{Extraction of the cross section}
\label{corr}
The events selected by the criteria described in section 2  are  used to determine the cross section $d \sigma / d \xpom (\gamma P \rightarrow XY)$ in the kinematic range defined in section 1. The HERWIG Monte Carlo, including BFKL pomeron exchange, is used to correct for losses and migration effects in $\xpom$, $\ypom$ and $t$. In the BFKL formalism at leading order, it does not make sense to run the coupling, and therefore $\alpha_s$ is fixed in the HERWIG generation at $0.17$. This corresponds at leading order to a hard pomeron intercept of $1+\omega_0$, where $\omega_0 = 0.45$.

The dominant contribution to the statistical error comes from the limited number of data events in the sample.
Systematic uncertainties are calculated on a bin by bin basis, and added in quadrature. The dominant error is due to the limited number of data events available to calculate the trigger efficiency, contributing a systematic error of approximately $30\%$ in each bin.

\section{Results}

\begin{figure} 
\centerline{\epsfig{file=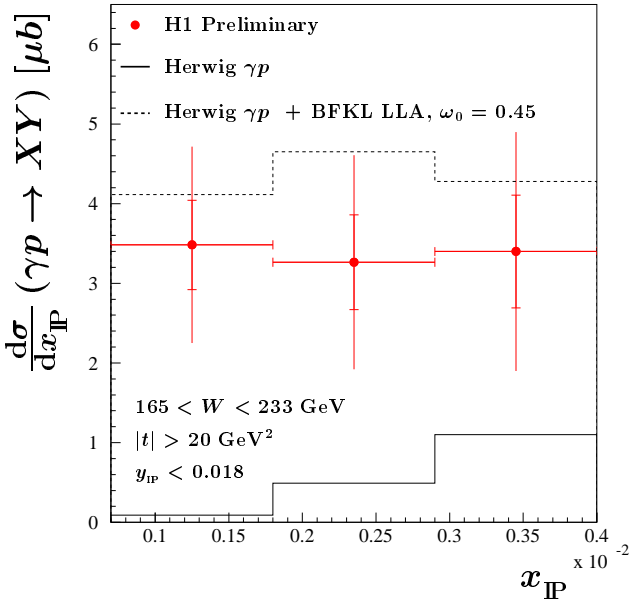,height=9.5cm}}
\caption{The differential cross section d$ \sigma / $d$ \xpom (\gamma P \rightarrow XY)$}
\label{xpom_fixW}
\end{figure}

The $\xpom$ distribution, corrected for detector effects, is shown in figure \ref{xpom_fixW}. The inner error bars are statistical and the outer error bars are the quadratic sum of the statistical and systematic errors. The solid line is the prediction from the HERWIG generator for all non-singlet exchange photoproduction processes. A significant excess above the expectation from the standard photoproduction model is observed. The dashed line shows the HERWIG prediction with the LLA BFKL prediction added. Good agreement is observed in both normalisation and shape. Care must be taken, however, in the interpretation of this result. There is a large theoretical uncertainty in the overall normalisation of the LLA BFKL cross section prediction. The agreement in normalisation may well therefore be fortuitous. It should also be noted that the shape of the $\xpom$ distribution in this region of phase space is not only determined by the underlying dynamics of the interaction, but also by kinematic effects. There is a kinematic limit on the lowest possible value of $\xpom$, set by the requirement that $|t|>20 \rm{GeV}^2$ and $165 < W < 233$ GeV, of $\xpom \simeq 7$ x $10^{-4}$ (see equation (3)). This forces the cross section down in the lowest $\xpom$ bin. The good agreement in shape with the BFKL Monte Carlo prediction, however, implies that the data are consistent with a value of $\omega_{0} \simeq 0.45$ within this model. 
Despite these limitations, however, with higher statistics the outlook for the future is promising. This measurement demonstrates that it is possible to extend greatly the reach in rapidity allowed by the gaps between jets approach. With the improved statistics already collected in the 1997 HERA running period, and higher luminosity in the future, a much more precise determination of the dependence of the cross section on $\xpom$, i.e. the energy dependence, will be possible.

\end{document}